\documentstyle[12pt,aps,psfig]{revtex}
\newcommand{\be}{\begin{equation}}
\newcommand{\ee}{\end{equation}}   
\newcommand{\bea}{\begin{eqnarray}}
\newcommand{\eea}{\end{eqnarray}}
\renewcommand{\d}{{\rm d}}
\newcommand{\gton}{\stackrel{>}{\sim}}
\newcommand{\lton}{\mathrel{\lower.9ex
                  \hbox{$\stackrel{\displaystyle <}{\sim}$}}}

\begin{document}
\title{Nuclear Broadening Effects on Hard Prompt Photons
at Relativistic Energies}
\author{A. Dumitru${}^a$, L. Frankfurt${}^{b,c}$, L. Gerland${}^b$, 
H. St\"ocker${}^b$, M. Strikman${}^d$}
\address{
{\it a} Physics Department, Columbia University, 538 West 120th Street,
New York, NY10027, USA\\
{\it b} Institut f\"ur Theoretische Physik der 
J.W.Goethe-Universit\"at\\
Robert-Mayer-Str. 8-10, D-60054 Frankfurt a.M., Germany
\\
{\it c} Tel Aviv University, Tel Aviv, Israel
\\
{\it d} Department of Physics, Pennsylvania State University,
University Park, PA 16802, USA}

\date{\today}

\maketitle
\begin{abstract}
We calculate prompt photon production in high-energy nuclear collisions.
We focus on the broadening of the intrinsic transverse momenta of the
partons in the initial state from nuclear effects, and their influence on
the prompt photon $p_t$ distribution. Comparing to WA98 data from Pb+Pb
collisions at $\sqrt{s}=17.4A$~GeV we find evidence for the presence of nuclear
broadening at high $p_t$ in this hard process. Below $p_t\sim2.7$~GeV
the photon distribution is due to small momentum transfer processes.
At RHIC energy, $\sqrt{s}=200A$~GeV,
the effect of intrinsic transverse momentum on the spectrum of
prompt photons is less prominent. The region $p_t=3-4$~GeV would be the
most promising for studying the nuclear broadening effects at that energy.
Below $p_t=2-3$~GeV the contribution from large momentum transfers flattens
out, and we expect that region to be dominated by soft contributions.
\end{abstract}

\section{introduction}
The WA98 collaboration recently reported data on direct photon production
in lead on lead collisions at CERN SPS energies
($\sqrt{s}=17.4A$~GeV) at $p_t\sim 1.5-4$~GeV~\cite{wa98}.
They made the interesting observation that in central Pb+Pb collisions
the multiplicity of direct photons {\em per nucleon-nucleon collision}
in that range of $p_t$ is enhanced relative to proton-proton collisions at
similar energy. (For a review of data and calculations of single photon 
production in proton and pion induced scattering off various targets
see~\cite{vogelsang}.)
It is the purpose of this paper to analyze to what extent,
and in which range of $p_t$, that data can be understood within
perturbative QCD (pQCD), after introducing an intrinsic transverse
momentum of the partons. That transverse momentum is due 
to the partons being confined in the initial-state nucleons, 
gluon bremsstrahlung,
as well as from multiple soft scattering of the nucleons prior to the 
hard scattering. It can be identified in the nuclear dependence of the
intrinsic 
transverse momentum of the partons predicted in~\cite{bodwin} and observed
recently for dimuon production in $p+A$ collisions~\cite{peng}.

Thus, the purpose of this paper is to analyse the direct photon data of
ref.~\cite{wa98} to probe nuclear broadening effects
on the transverse momentum distribution of direct photons.
Those effects are expected to increase the prompt photon cross section
strongly, because a part of the $p_t$ of the photon can be ``supplied''
by the incoming partons rather than in the elementary semi-hard scattering
itself~\cite{owens}. 

Even so, we explicitly restrict the computation to include only the
contribution from parton-parton scattering amplitudes at a minimum momentum 
transfer of $\ge 1$~GeV. Thus, comparing the calculated $p_t$ distribution 
of photons to the data, one can determine the $p_t$ scale below which
production of single photons
is dominated by reactions at small momentum transfer, $<1$~GeV. That is
probably the region where a pQCD description looks suspicious.
Note that without imposing a cut on the
minimum momentum transfer in elementary parton-parton scatterings the
perturbative contribution becomes very large at low $p_t$,
precisely because of the fact
that $p_t\sim1-2$~GeV photons can be produced with almost no momentum
transfer, the $p_t$ being supplied by the intrinsic $k_t$ of the partons.
Therefore, care must be taken when comparing perturbative
computations accounting for the violation of the DGLAP approximation 
due to parton intrinsic transverse momentum to data.
This primary transverse momentum is to large extent due to the gluon 
bremsstrahlung. With increasing energy, primary transverse
momentum plays a less important role because a lot of gluons 
are radiated ``long before'' the hard collision~\cite{collins}.
Thus the distribution of partons near the point of the hard collision will be
determined mostly by gluon bremstrahlung, which is universal -- independent of
the target.

We shall also make predictions for Au+Au collisions at nominal BNL-RHIC
energy, $\sqrt{s}=200A$~GeV. Prompt photon spectra at RHIC will be measured
in the future by the PHENIX collaboration. We shall discuss the importance of
intrinsic transverse momentum at that energy as well, and identify
the $p_t$ where the photon distribution from pQCD interactions with momentum
transfer $\ge1$~GeV flattens out. 

The significant effects from nuclear broadening of the transverse parton
distribution on photon production in $A+A$ collisions are in 
general agreement with the description of the yield of electrically neutral 
pions in~\cite{MGPL}. It proved essential for the understanding of
$\pi^0$ production in Pb+Pb at SPS energy~\cite{wa98pi0}
($\sqrt{s}=17.4A$~GeV; $p_t=1-4$~GeV) to account for the $p_t$-broadening.
Also, measurements of both the prompt photon and pion production
cross section in future $p+A$ experiments at BNL-RHIC can determine the
scaling of the nuclear broadening with $A$~\cite{MGPL,plf};
that could provide additional information to that obtained from
the Drell-Yan process and $J/\Psi$, $\Upsilon$ production.

\section{Prompt Photon $p_t$ distribution in $p+p$}\label{ppgamma}

We consider the contributions from Compton-like scattering
($g+\stackrel{(-)}{q}\rightarrow \gamma+\stackrel{(-)}{q}$), annihilation
($q+\overline{q}\rightarrow \gamma+g$), plus collinear bremsstrahlung off
a (anti-)quark produced at midrapidity.
Assuming the applicability of the QCD factorization theorem, the 
corresponding expressions for the Compton and annihilation
subprocesses in $p+p\rightarrow\gamma+X$ are~\cite{owens}
\be \label{ppXsec}
E\frac{\d\sigma_\gamma}{\d^3p} = \sum \int \d x_a \d x_b \d^2k_{ta} \d^2k_{tb}
f(k_{ta}) f(k_{tb}) 
G_{a/A}(x_a,Q^2)
G_{b/B}(x_b,Q^2) \frac{\hat{s}}{\pi} \frac{\d\sigma}{\d\hat{t}}\delta
(\hat{s}+\hat{t}+\hat{u}) \Theta_0~.
\ee
$\hat{s}$, $\hat{t}$, $\hat{u}$ denote the Mandelstam variables for the
$a+b\rightarrow\gamma+c$ elementary process.
The sum extends over all possible partons in the initial state, i.e.\
gluons and $u$, $d$, $s$ (anti-)quarks, and over 
any final state interactions.
${\d\sigma}/{\d\hat{t}}$ denotes
the elementary hard-scattering cross section for the corresponding
process, averaged over all possible spin and color orientations in the
initial state, and summed over those in the final state.
Explicit expressions can be found e.g.\ in~\cite{owens}. 

The $G_{a/A}(x_a,Q^2)$ are the twist-2 infrared dominated matrix elements,
for which we employ the CTEQ4L parametrization~\cite{lai1}.
(We consider only the processes with unpolarized particles, 
so twist-4 would be the next
to contribute.) At rather small-$x$, nuclear 
shadowing of the parton distribution functions
may result in considerable
suppression of the gluon and sea-quark distributions at moderate
values of $Q^2$~\cite{fs88}. This may then in turn affect the $p_t$
distribution of prompt photons at collider energies (RHIC, LHC) in the
$p_t$-range of a few GeV~\cite{hamdum,jamal}.
On the other hand, at CERN-SPS energy,
$\sqrt{s}\simeq 17-20A$~GeV, and for $p_t\sim1$~GeV, we rather deal with
the parton distributions at $x\sim0.1$, which is in the domain of 
{\em antishadowing} effects. The antishadowing effect is rather small, however:
EKS~\cite{EKS} and
FGS~\cite{FGS} estimates predict about $10\%-15\%$ enhancement.
As we shall see below, this appears to be well within the uncertainties
of the value of the intrinsic transverse momentum, and thus
will not be considered here in more detail.

The functions $f(k_t)$ parametrize the transverse momentum of the partons
in the initial state. For simplicity, we assume a Gaussian distribution,
\be \label{Gaussipt}
f(k_t) = \frac{1}{\pi \langle k_t^2\rangle} e^{-k_t^2/\langle k_t^2\rangle}~.
\ee
The distribution~(\ref{Gaussipt})
contains one %freely 
tunable parameter, namely the average
intrinsic transverse momentum of the partons in the initial state,
$\langle k_t^2\rangle$. We shall present results for various values of
$\langle k_t^2\rangle$ below. The limit $\langle k_t^2\rangle\rightarrow0$
recovers the usual collinear factorization.

The intrinsic $k_t$ for plane waves bound in a nucleon can be estimated
by applying the uncertainty principle:
\be
\sqrt{k_t^2} \approx {\pi\over 2r_N}  \approx 0.37~\mbox{GeV}~.
\ee
Here, $r_N \approx 0.85$~fm is the radius of a nucleon.
Initial state gluon bremsstrahlung can give rise to larger intrinsic $k_t$
on the order of 1~GeV.

For the factorization scale we assume $Q^2=(2p_t)^2$, and as we restrict
ourselves to the leading logarithmic approximation, we assume that the
renormalization scale is the same. We employ the one-loop expression for
the running coupling constant $\alpha_s(Q^2)$ for $N_f=4$, and $\Lambda_{\rm
QCD}=236$~MeV, as appropriate for CTEQ4L. Collinear divergencies
arising from partonic processes with large intrinsic transverse momentum
are cut off by
\be \label{Theta_tot}
\Theta_0 = \Theta_1 \Theta_2 \,\,\Theta\left(Q_c^2-1{\rm GeV}^2\right)~,
\ee
where we define 
\be
Q^2_c = \frac{2\hat{s}\hat{t}\hat{u}}{\hat{s}^2+\hat{t}^2+\hat{u}^2}~.
\ee
In this way, only subprocesses with momentum transfer $\ge1$~GeV are
taken into account, for which one can hope that factorization is a reasonable
assumption. The functions $\Theta_1$ and $\Theta_2$ ensure that
parton $a$ is moving to the right, while parton $b$ is moving to the left,
\be \label{Theta_12}
\Theta_1 = \Theta(k_{za})\quad,\quad
\Theta_2 = \Theta(-k_{zb})~.
\ee
Using on-shell kinematics, the four-momenta of the incoming
partons are given by
\bea
(E_a,\vec{k}_{ta},k_{za}) &=& \left( \frac{1}{2}x_a\sqrt{s}+
   \frac{{k}^2_{ta}}{2x_a\sqrt{s}},\vec{k}_{ta},
   \frac{1}{2}x_a\sqrt{s}-\frac{{k}^2_{ta}}{2x_a\sqrt{s}}\right)~, \\
(E_b,\vec{k}_{tb},k_{zb}) &=& \left( \frac{1}{2}x_b\sqrt{s}+
   \frac{{k}^2_{tb}}{2x_b\sqrt{s}},\vec{k}_{tb},
   -\frac{1}{2}x_b\sqrt{s}+\frac{{k}^2_{tb}}{2x_b\sqrt{s}}\right)~.
\eea
After using the $\delta$-function in eq.~(\ref{ppXsec}) to solve for
$x_b$, the remaining integrals can be performed numerically via Monte-Carlo
techniques. Numerical convergence can be checked by comparing
spectra obtained with varying sequences of random numbers and increasing
the number of Monte-Carlo points.

Collinear bremsstrahlung is computed as follows. First, the cross-section
for (anti-)quark production at $y_{cm}=0$ with transverse momentum $p_{tc}$
is computed in the standard fashion, as described above. This is then
convoluted with the QED collinear fragmentation function $z_cD_{\gamma/q}(z_c)=
{\alpha}[1+(1-z_c)^2]\log(p_t^2/\Lambda^2_{\rm QCD})/{2\pi}$
into $\gamma+X$, where $z_c=p_t/p_{tc}$,
\bea 
E\frac{\d\sigma_\gamma}{\d^3p} &=& \sum \int \d x_a \d x_b \d z_c
\d^2k_{ta} \d^2k_{tb}\nonumber \\
& & f(k_{ta}) f(k_{tb}) 
G_{a/A}(x_a,Q^2)
G_{b/B}(x_b,Q^2) 
D_{\gamma/c}(z_c)
\frac{\hat{s}}{\pi z_c^2} \frac{\d\sigma}{\d\hat{t}}\delta
(\hat{s}+\hat{t}+\hat{u}) \Theta_0~. \label{ppbremsXsec}
\eea
Here, we employ $Q^2 = (2p_{tc})^2$ for the factorization and renormalization
scale of the hard process. For the bremsstrahlung contribution we do not
sum explicitly over all hard subprocesses leading to the production of a
(anti-)quark at midrapidity but rather employ the method of effective
structure function,
\be
G_{a/A}(x_a,Q^2)G_{b/B}(x_b,Q^2)
 = \left(g(x_b,Q^2) + \frac{4}{9}\sum_{f} q_f(x_b,Q^2)\right)
   \sum_{f} e_f^2q_f(x_a,Q^2)~,
\ee
where $e_f$ is the fractional electric charge of the
splitting (anti-)quark. Further, ${\d\sigma}/{\d\hat{t}}=
\pi(\alpha_s/\hat{s})^2 (\hat{s}^2+\hat{u}^2)/\hat{t}^2$. The
$\Theta$-function is defined as in~(\ref{Theta_tot},\ref{Theta_12}).
Note that in these expressions $\hat{s}$, $\hat{t}$, $\hat{u}$ refer to
the partonic subprocess, ``before'' splitting of the final-state (anti-)quark
into $\gamma+X$.

It is common practice to multiply the r.h.s.\ of~(\ref{ppXsec})
and~(\ref{ppbremsXsec}) by a $K$-factor in order to account for NLO
contributions. Full NLO computations have been performed~\cite{NLO}
 for the case
without intrinsic $k_t$ of the partons and indicate $K\simeq1.8-2.0$ for
$p+p$ collisions at energies around $\sqrt{s}=20$~GeV, and for
photon transverse momenta in the range $2-4$~GeV.
Our major interest is in the relative yields
of prompt photons in $p+p$ and in $A+A$ collisions, and in that sense
the $K$-factor plays no essential role.
In any case, we shall fix the value of $K$ from $p+p$ and then keep its
value for $A+A$.

\begin{figure}
\centerline{\psfig{figure=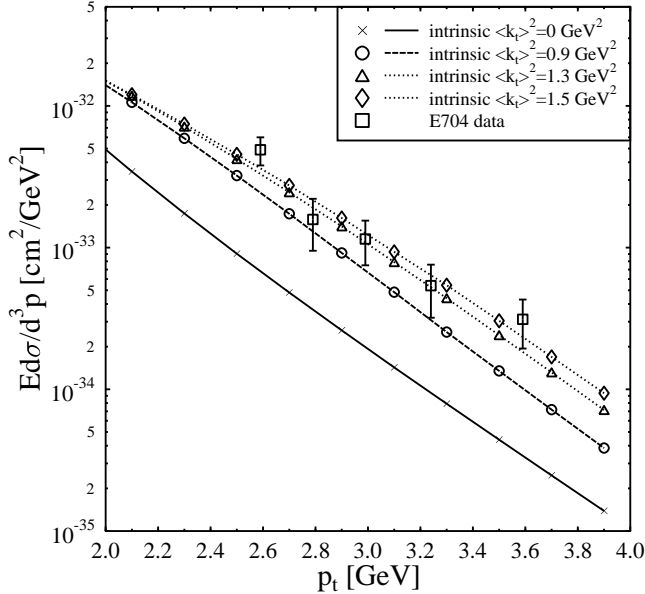,height=10cm}}
\caption{The differential direct photon cross section vs.\ $p_t$ 
(at center-of-mass rapidity) for
proton-proton collsions at $\sqrt{s}=19.4$~GeV.}
\label{fig_pp194}
\end{figure}
In Fig.~\ref{fig_pp194} we compare our results for $p+p$ at
$\sqrt{s}=19.4$~GeV, for various intrinsic $\langle k_t^2\rangle$ and
$K=2$, to E704 data~\cite{E704}.  Clearly, the
calculation assuming pure DGLAP evolution, i.e.\ $\langle k_t^2\rangle=0$,
falls short by almost an order of magnitude. 
The best fit is obtained using $\langle k_t^2\rangle=1.3-1.5$~GeV$^2$.
At first sight this appears somewhat larger than
the ``standard'' $\langle k_t^2\rangle=0.9$~GeV$^2$ from~\cite{owens,w+w}.
However, recall that we employ a larger $Q^2=4p_t^2$ scale
than~\cite{owens,w+w}, who used $Q^2=p_t^2/2$.
In accord with the picture that the intrinsic transverse momentum is
due mainly to gluon radiation in the initial state, a larger $Q^2$ scale
therefore has to correspond to larger $\langle k_t^2\rangle$.
This general expectation is confirmed by our results.
(The value for $\langle k_t^2\rangle$ therefore only acquires physical meaning
together with a definition of $Q^2$.)

The above calculation serves to fix the $K$-factor and the strength of the
DGLAP violation on the $p+p$ level. From there we conclude that with
$K=2$, $Q^2=4p_t^2$, a value of $\langle k_t^2\rangle\simeq1.3-1.5$~GeV$^2$
is required to fit the data at that energy.

\section{Hard Photon Production in Nucleus-Nucleus collisions}
\label{gamma_AA}
In the previous section we discussed hard photon production
in $p+p$ or $N+N$ collisions (we neglect the differences 
between protons and nucleons in our calculations).
To extend our calculations to nucleus-nucleus collisions we 
have to substitute the parton distribution functions $G_N(x,Q)$ in 
eq.~(\ref{ppXsec}) by $G_A(x,Q,\bbox{b}')$ for the projectile and 
$G_B(x,Q,\bbox{b}-\bbox{b}')$ for the target nucleus. Here, 
$G_A(x,Q,\bbox{b}')=A\cdot T_A(\bbox{b}')G_N(x,Q)$ and 
$G_B(x,Q,\bbox{b}-\bbox{b}')=B\cdot T_B(\bbox{b}-\bbox{b}')
G_N(x,Q)$~\cite{liuti}, with $\bbox{b}$
the nucleus-nucleus impact parameter, and $\bbox{b}'$ the distance from the 
center of the projectile. Note that $\bbox{b}$ and $\bbox{b}'$
are two-dimensional vectors.

The nuclear thickness functions 
$T_{A,B}(\bbox{b}')$ are defined as
\begin{equation}
T_{A,B}(\bbox{b}')=\int{\rm d}z \,\rho_{A,B}(\bbox{b}',z)
\mbox{ with } \int {\rm d}^2\bbox{b}'
{\rm d}z \, \rho(\bbox{b}',z)=1\;.
\label{thickness}
\end{equation}
$A\; (B)$ are the number of nucleons in the projectile (target).
We account for the dependence on the
nucleus nucleus impact parameter $b$ as due to the geometry of the collision 
only. If one integrates over all impact parameters to compare with minimum bias
data, this yields
\begin{equation}
\sigma_{\gamma}^{AB}=AB\cdot \sigma_{\gamma}^{NN}\;.
\end{equation}

To compare to the experimental data by the WA98 collaboration
we need to calculate the direct photon cross section for their
centrality trigger. To take into account the dependence
of the nuclear parton distribution functions on the geometry we employ
the semiclassical approximation of the Glauber-model that was developed
for hadron nucleus collisions in ref.~\cite{glauber} and extented in ref.~\cite{czyz}
to nucleus nucleus collisions.

The inclusive spectrum of hard photons in  
$A+B$ collisions at the impact parameter $\bbox{b}$ is
\begin{equation}
E_\gamma {{\rm d} N_{\gamma}^{A B}\over 
{\rm d}^3 \bbox{p}_\gamma}(\bbox{b}) =
AB\cdot T_{AB}(\bbox{b})\;E_\gamma {{\rm d} 
\sigma_\gamma^{NN}\over {\rm d}^3 \bbox{p}_\gamma} \;.
\label{ww47}    
\end{equation}
$E_\gamma$ and $\bbox{p}_\gamma$ are the energy and the momentum of the
emitted photon, and
$\sigma_\gamma$ is the inclusive cross section for direct photon production.

$T_{AB}$ is the nucleus-nucleus thickness function. For large nuclei with 
a slowly varying density distribution $T_{A B}$ is given by
\begin{equation}
T_{AB}(\bbox{b})=\int{\rm d}^2\bbox{b}' \, T_A(\bbox{b}')T_B(\bbox{b}-\bbox{b}')\;.
\label{ww50}
\end{equation}
In a $A+B$ collision at impact parameter $\bbox{b}$, the nucleus-nucleus
thickness function allows only collisions of partons in overlapping slabs,
i.e.\ of partons with the
same coordinates in the plane transverse to the beam direction.

Finally, the differential number of direct photons in nucleus-nucleus 
collisions integrated over impact parameters $0<b<b_m$ is given by the ratio of
the differential cross section to the inelastic nucleus-nucleus cross section,
both being integrated over the region $[0,b_m]$:
\begin{equation}
E_\gamma {{\rm d}  N_\gamma^{AB}\over {\rm d}^3 {\bf p}_\gamma}(b_m)=
{E_\gamma \int_0^{b_m} {\rm d}^2\bbox{b}{{\rm d} \sigma_{\gamma}^{A B}\over
{\rm d}^3 {\bf p}_\gamma}(\bbox{b})\over \sigma_{inel}^{AB}(b_m)}=
{E_\gamma{{\rm d}\sigma^{\gamma}_{pp}\over {\rm d}^3 {\bf p}_\gamma}
\int_0^{b_m} {\rm d}^2\bbox{b}AB\cdot T_{A B}(\bbox{b}) \over
\int_0^{b_m} {\rm d}^2\bbox{b} (1-(1-T_{AB}(\bbox{b})\sigma_{in}^{NN})^{AB})}~.
\label{ww54}
\end{equation}
{}From this equation, we can also obtain the prompt photon multiplicity
for $p+A$ scattering. To that end, the nucleus-nucleus thickness function
$T_{A B}(\bbox{b})$ has to be replaced by the nucleus thickness function
$T_{A}(\bbox{b})$, and $B=1$ throughout the equation. For $p+A$ we have
$b_{m}=\infty$, corresponding to minimum bias scattering.

In our calculations we used the Woods-Saxon density distribution of 
ref.~\cite{devries} and assumed $\sigma_{in}^{NN}=30$ mb at $\sqrt{s}
\sim20A$~GeV. That yields a
total inelastic cross section for Pb+Pb of $\sigma_{tot}^{PbPb}=7100$~mb.
$\sigma_{tot}^{PbPb}$ is given by the denominator of eq.~(\ref{ww54}) with
$b_m=\infty $. For large values of $AB$ one can approximate the
denominator by 
\begin{equation}
\int {\rm d}^2\bbox{b}\; (1-(1-T_{AB}(\bbox{b})\sigma_{in}^{NN})^{AB})
\approx\int {\rm d}^2\bbox{b}\; (1-\exp(-ABT_{AB}(\bbox{b})\sigma_{in}^{NN}))\;.
\end{equation}
In ref.~\cite{czyz} this is called optical limit for the inelastic cross
section.
 
The geometrical
factor in eq.~(\ref{ww54}) which multiplies the $p+p$ differential
cross section to yield the photon multiplicity equals 24.3/mb for
the 10\% most central Pb+Pb collisions, as analysed by WA98 (corresponding to
$b_m\approx 4.8$ fm in our calculation).
For minimum bias $p+$C it is 0.056/mb, and 0.127/mb for $p+$Pb.
These values allow to calculate the yield of prompt photons in $p+A$
collisions at SPS energy from the cross section in $p+p$ as shown in   
Fig.~\ref{fig_pp194}.

For nominal BNL-RHIC energy, $\sqrt{s}=200A$~GeV, using $\sigma_{in}^{NN}=
40$~mb, we obtain for that factor 0.119/mb for $p+$Au, 0.048/mb for $p+$C,
0.062/mb for $p+$S, and finally 23.1/mb for Au+Au (with 10\% centrality cut;
$\sigma_{tot}^{\rm AuAu}=7024$~mb; $b_m\approx 4.7$~fm).

\section{Nuclear broadening of the intrinsic $k_t$ distribution}

The nuclear broadening of the intrinsic transverse momentum distributions of the
partons was observed experimentally by the broadening of the transverse
momentum distribution of final state particles. For example, ref.~\cite{Cronin}
analyzed $p_t$ distributions of mesons made of light quarks
($\pi$, K, $\dots$); ref.~\cite{peng} analyzed
Drell-Yan pairs and quarkonium states ($J/\psi$ and $\Upsilon$).

The aim of this paper is to discuss whether that broadening can
explain the discrepancy of the WA98 data for Pb+Pb collisions to
simple extrapolations of proton-proton 
data and calculations. We define the broadening as
\begin{equation}
\Delta k_t^2\equiv \langle k_t^2 \rangle_{AB}-\langle k_t^2 \rangle_{pp}~,
\end{equation}
where $\langle k_t^2 \rangle_{AB(pp)}$ is the average transverse
momentum in $A+B$ ($p+p$) collisions. For Pb+Pb collisions values of 
$\Delta k_t^2=0-1$~GeV${}^2$ will be used in our
calculations below. Larger broadening, as suggested for example by
ref.~\cite{baier} for hot QCD matter, seem to be ruled out at SPS energies as
will be shown in section~\ref{result}. The value suggested in~\cite{baier} for 
cold (hot) QCD matter is
\begin{equation}
\Delta k_t^2=0.2\;(3.0)\mbox{ GeV}^2\cdot{L\over 10\;\rm{fm}}
\label{deltap}
\end{equation}
$L$ is the length of the QCD medium through which the parton propagates
before the hard process.
The average thickness of a sphere is the volume divided by the area:
\begin{equation}
\langle L\rangle_{sphere}=({4\over 3}\pi R^3)/(\pi R^2)={4\over 3}R_{A,B}~.
\end{equation}
For proton-nucleus collisions that value has to be divided by two, since we 
are interested in initial state 
interaction only (the outgoing photon does not interact strongly).
For nucleus-nucleus collisions another factor two is necessary, since both
the intrinsic transverse momentum of the projectile and of the target can 
increase due to soft inital-state interactions. 
$R_{A,B}$ denotes the radius of the projectile (target) nucleus.
For Pb+Pb this means $L\simeq 2\cdot {2\over 3}R={4\over 3}\cdot 
6.6\mbox{ fm}=8.8\mbox{ fm}$. From eq.~(\ref{deltap}) we thus obtain for
cold (hot) QCD matter
\begin{equation}
\Delta k_t^2=0.2\;(3.0)\mbox{ GeV}^2\cdot 0.88=0.176\;(2.64)\mbox{ GeV}^2.
\label{deltap2}
\end{equation}
For gluons $\Delta k_t^2$ should be larger by the
ratio of the Casimir operators of the octet and triplet 
representations of color-SU(3), $=9/4$.
While these number are only rough estimates of the order of magnitude, we
shall see
below that $\Delta k_t^2\sim 0.5-1$~GeV$^2$ is in the range needed to
explain the WA98 prompt photon excess for $p_t\gton2.5$~GeV.
 
Such small broadening is
also compatible with the fits to $p+A$ data at Fermilab energies~\cite{peng}.
Those fits yield
\begin{equation}
\Delta k_t^2=0.133\;(0.027)\mbox{ GeV}^2
\left(\left({A\over 2}\right)^{1\over 3}-1\right)
\label{pengfits}
\end{equation}
for $J/\psi$ (Drell-Yan pair) production. For $p+$Pb this is
$\Delta k_t^2=0.49\;(0.10)\mbox{ GeV}^2$. 
Assuming linear scaling~\cite{baier} of $\Delta k_t^2$ with
the length traversed through the QCD matter, $L$,
the above value has to be multiplied by two, for minimum bias lead on lead 
collisions:
\begin{equation}
\Delta k_t^2({\rm Pb+Pb})=0.98\;(0.2)\mbox{ GeV}^2.
\label{pengfitsresult}
\end{equation}
Regarding the Drell-Yan process, the $\Delta k_t^2$ broadening of the
quarks in the initial state equals that of the muon pair in the final
state. In our case, however, the dominant processes involve a
gluon in the initial state, and so the broadening in the initial
state is expected to be larger by a factor $(1+9/4)/(1+1)$, i.e.\
$0.2$~GeV$^2\rightarrow0.325$~GeV$^2$.
For a central trigger, $L$ and thus $\Delta k_t^2$ could even
be somewhat larger.

To summarize, values for $\Delta k_t^2$ up to 1~GeV$^2$
appear to be within the range given by various fits to data,
and shall be employed below for the computation of prompt photon
production in nuclear collisions.

\section{Results for CERN-SPS energy}\label{result}

\begin{figure}
\centerline{\psfig{figure=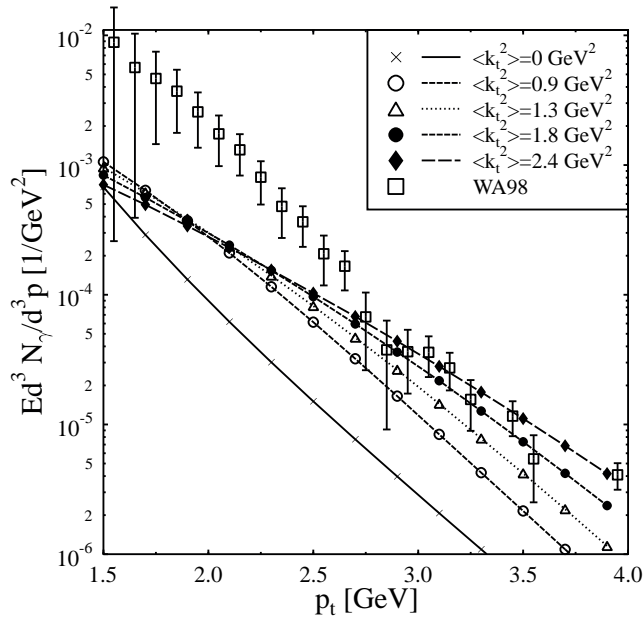,height=10cm}}
\caption{The transverse momentum distribution of prompt photons at
center-of-mass rapidity
in the 10\% most central Pb+Pb reactions at $\sqrt{s}=17.4A$~GeV.
The data~\protect\cite{wa98} is compared to calculations for
various intrinsic $\langle k_t^2\rangle$.
\label{photon3}}
\end{figure}
Fig.~\ref{photon3} depicts the differential multiplicity of direct photons vs.\
transverse momentum in central Pb+Pb reactions at $\sqrt{s}=17.4A$~GeV.
We compare calculations with various values for the
intrinsic transverse momentum, $\langle k_t^2\rangle$, all corresponding
to $K=2$. The data are from ref.~\cite{wa98}.
Again, as was the case for $p+p$ collisions at similar energy,
the calculation assuming pure DGLAP evolution underestimates
the data strongly. However, the ``standard'' intrinsic transverse
parton momentum of $\langle k_t^2\rangle\simeq1.3$~GeV$^2$ extracted
already on the $p+p$ level (see section~\ref{ppgamma}) improves the
agreement a lot.
Within the experimental error bars, the WA98 data leaves room for up to
$\Delta k_t^2\sim1$~GeV$^2$ additional broadening from nuclear effects.
However, we can also conclude that transverse momentum broadening much beyond
$\Delta k_t^2\simeq1$~GeV$^2$ is not seen in the WA98 data.

Another interesting result is that the calculated photon spectrum below
$p_t\lton2.5$~GeV depends only weakly on the amount of intrinsic transverse
momentum. That is because of the cut-off on the momentum transfer,
eq.~(\ref{Theta_tot}). Therefore, the excess photons seen in the data, as
compared to our calculation, can be attributed to small-momentum transfer
scatterings (final-state interactions). The computation of that
contribution to the photon yield is, however, out of the scope of the
present manuscript; see e.g.~\cite{Gallmeister:2000si} for recent work,
and references therein.

\section{Results for BNL-RHIC energy}\label{resultRHIC}
We now turn to nuclear collisions at collider energy. Here, we focus on
Au+Au collisions at $\sqrt{s}=200A$~GeV and central rapidity, $y\approx0$.
(Estimates for $p+A$ collisions can be obtained from our Au+Au results 
by an appropriate scaling as explained in section~\ref{gamma_AA}.)
At relatively small transverse momenta nuclear shadowing effects~\cite{fs88}
may reduce the prompt photon cross section somewhat~\cite{hamdum,jamal}.
However, for $p_t\gton2$~GeV the effect is not very large, $\lton20\%$,
and we shall not investigate it in more detail here.
On the other hand, it has been suggested that at high energies
the gluon distribution in nuclei may saturate at small fractional light-cone
momentum $x$~\cite{MV} (but much larger $x$ than for protons).
This must happen at least when the unitarity bound for the inelastic
scattering of a QCD dipole off a nucleus is reached in the $x-Q^2$
plane~\cite{FGS},
perhaps earlier. Simply speaking, the nucleus becomes a ``black disc''
in the infinite momentum frame.
If heavy nuclei but not protons become black, or almost black,
jets with $p_t<Q_{black}$ will be absorbed in the final state
in $A+A$ scattering\footnote{They might produce
a ``saturated plasma''~\cite{satplasma} at central rapidity which
would then also contribute to photon production~\cite{Gallmeister:2000si}.},
but not in $p+p$ scattering. So
high $p_t$ jets will be expected for $p_t>Q_{black}$
only. Thus, qualitatively, the $p_t$ distribution should be rather different
{}from $p+p$ at not too large $p_t$. The yield of jets will be suppressed but
the transverse momentum distribution will be wider.

For $A=200$ nuclei, the unitarity limit for the inelastic $q\bar{q}$-nucleus
cross section is reached at $x\sim10^{-2}$ for $Q\sim1.5$~GeV,
and at $x\sim10^{-3}$ for $Q\sim2.5$~GeV
(if nuclear shadowing is weak)~\cite{FGS}. If shadowing is also taken into
account, the $x$-values where the unitarity limit is approached become even
smaller~\cite{FGS}.
Therefore, at rapidity $y\sim0$ and at RHIC energy, $\sqrt{s}=200A$~GeV,
we expect only small effects on prompt photon production.

At high energy, the bremsstrahlung contribution to the prompt photon yield
is substantial~\cite{hamdum,jamal,NLO}. In this process, a (anti-)quark  
is produced first in the central rapidity region with transverse momentum
$p_{tc}$ (see section~\ref{ppgamma}).
That quark then radiates an almost collinear   
photon with transverse momentum $p_t=z p_{tc}$.
In principle, the quark could in fact suffer further collisions which in turn 
may suppress the contribution from soft (i.e., small $z$) bremsstrahlung to 
prompt photon production~\cite{jamal,hamdum2}. This is due to the well known
Landau-Pomeranchuck-Migdal effect. The quark could also suffer some energy loss
prior to emitting the photon. The radiative energy
loss of the quark is mainly due to radiation of soft gluons.
Therefore, the formation time (or the formation length) of the radiated
gluons cannot be much smaller than that of the photons. Thus,
quark energy loss might not interfere with the radition of the photons in the
final state. In any case, the complete extinction of all
bremsstrahlung would represent an upper limit for the sum of both
effects~\cite{hamdum2}. This would change
our result by less than a factor of two. Therefore, we do not discuss  
these effects in more detail here (see~\cite{jamal}),
since their largest possible impact
on the photon production cross section is smaller than that from
nuclear broadening of the intrinsic transverse momentum.

\begin{figure}
\centerline{\psfig{figure=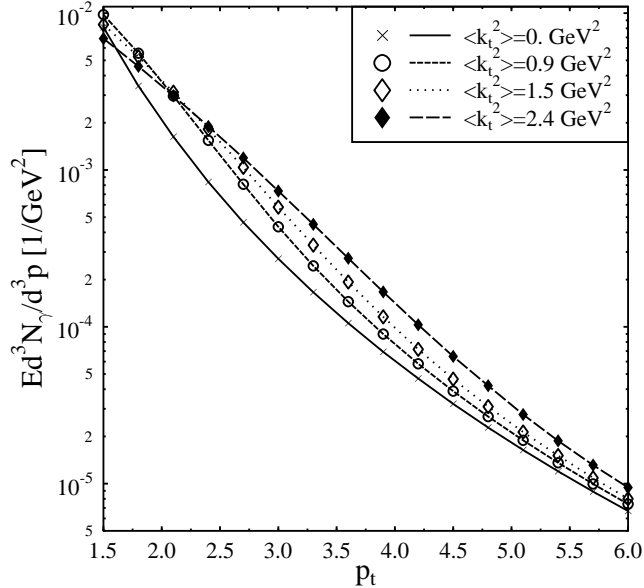,height=10cm}}
\caption{The transverse momentum distribution of prompt photons at
center-of-mass rapidity
in the 10\% most central Au+Au reactions at $\sqrt{s}=200A$~GeV for
various intrinsic $\langle k_t^2\rangle$.
\label{photon_rhic}}
\end{figure}
In Fig.~\ref{photon_rhic} we show prompt photon spectra at RHIC energy,
for the $10\%$ most central Au+Au events at $\sqrt{s}=200A$~GeV.
As before, we assumed $K=2$. One observes that the effect of intrinsic
transverse momentum in the initial state is much less prominent than at
lower energy, in agreement with the results of~\cite{MGPL} for
pion production. The reason for this behavior is that the photon spectrum
becomes ``harder'' at higher energy, i.e.\ it decreases less steeply with
$p_t$.
Nevertheless, at $p_t\sim3-4$~GeV the intrinsic parton transverse momentum
can increase the prompt photon multiplicity by up to a factor of 3.
That would be the most promising kinematical domain for an experimental
study of the nuclear broadening of the intrinsic transverse momentum
distributions. The region of smaller $p_t$ might be dominated by soft
contributions, as indicated by the flattening of the spectra in
Fig.~\ref{photon_rhic}. On the other hand, at larger $p_t$ intrinsic
transverse momentum becomes less effective.

\section{Summary and Conclusion}
Prompt photon data from $p+p$ scattering at $\sqrt{s}=19.4$~GeV
obtained by the E704 collaboration~\cite{E704} reveals that standard DGLAP
evolution without intrinsic transverse momentum, and for our
choice for the factorization scale $Q=2p_t$, underestimates
that data. Including an intrinsic transverse momentum of the initial-state
partons of $\sim1.3$~GeV$^2$ allows for a fair description
of the E704 data.

Further, we presented results for prompt photon $p_t$ distributions in
nuclear collisions, assuming $A$-scaling as suggested by the geometry
of the Glauber incoherent approach. At SPS energy, a good description of the
data for $p_t\gton2.7$~GeV requires additional intrinsic $k_t$ of the
partons in the initial state, i.e.\ nuclear broadening of the transverse
parton distributions by $\Delta k_t^2 \sim 0.5-1$~GeV$^2$ (for the same
factorization scale as above). Thus, the WA98 data provide evidence
for the nuclear broadening effect in the
$p_t$ distribution of prompt photons. Photons appear more appropriate 
than hadrons to
test nuclear broadening of the intrinsic transverse momentum because they
do not suffer from rescattering in the final state.

Below $p_t\sim2.7$~GeV, both the yield and the slope of the WA98 photon
data is not described as being due to hard scatterings with momentum
transfer $\ge1$~GeV. At $p_t\sim2$~GeV, prompt photons from hard
collisions underestimate the WA98 data by almost an order of magnitude,
for any amount of intrinsic $\langle k_t^2\rangle$.
Thus, that kinematic domain is dominated by small momentum transfer processes.

At RHIC energy, the effect of the intrinsic transverse momentum is much less
prominent. Only at $p_t\sim3-4$~GeV can intrinsic parton transverse momentum
increase the prompt photon multiplicity by up to a factor of 3.
That would thus be the most promising kinematical domain for an experimental 
study of nuclear broadening of the intrinsic transverse momentum at RHIC.
Again, at $p_t\lton 2-3$~GeV the realistic prompt photon distribution 
(incl.\ intrinsic $k_t$) from large momentum transfer scattering flattens out,
that region being most likely dominated by soft contributions.
\acknowledgements 
We thank T.\ Peitzmann and M.\ Mora for discussions on the
WA98 data, and Y.~Dokshitzer, J.~Jalilian-Marian and I.\ Vitev
for discussions on energy loss and fragmentation time scales.
A.D.\ acknowledges support from DOE Research Grant DE-FG-02-93ER-40764.
M.S.\ thanks the US Department of Energy for financial support.
L.G.\ would like to thank the Josef Buchmann Foundation for 
financial support. H.S.\ was supported in parts by BMBF, DFG, and GSI.

\end{document}